\RequirePackage{lineno}
\documentclass[aps,prd,showpacs,amsmath,twocolumn]{revtex4}
\usepackage{graphicx}
\usepackage{epsfig}
\usepackage{dcolumn}
\usepackage{bm}
\usepackage{subfigure}
\usepackage[titletoc]{appendix}
\usepackage{verbatim}

\usepackage{amssymb}
\usepackage{amsfonts}
\usepackage{amsmath,array}
\usepackage{epsfig}
\usepackage{graphicx}
\usepackage{epstopdf} %eps2pdf
\usepackage{dcolumn}% Align table columns on decimal point
\usepackage{bm}% bold math
\usepackage{lineno}
\usepackage{multirow}
\usepackage{color}
\usepackage[perpage,symbol]{footmisc}%% widetext enviorenment
\usepackage{subfigure}
\setfnsymbol{wiley}

\hfuzz=\maxdimen
\include{def-com}
\def \zcs {Z_{cs}}

\def \ee {e^+e^-}

\def \dst {D^{*0}}

\begin{document}
%\linenumbers
%\preprint{} \preprint{\vbox{ \hbox{  }\hbox{Ver3.5} }}
\title{\boldmath Spin and polarization analysis of $\zcs$ state}
\author{Hong Chen$^a$, Qi Huang$^{b}$, Rong-Gang Ping$^{b,c}$\\
a) School of Physical Science and Technology, Southwest University, Chongqing 400715, China\\
b) University of Chinese Academy of Sciences (UCAS), Beijing 100049, China\\
c) Institute of High Energy Physics, Chinese Academy of Sciences,\\
 P.O. Box 918(1), Beijing 100049, China\\
\date{ }}
\email[]{pingrg@ihep.ac.cn, chenh@swu.edu.cn}
\begin{abstract}
A polarization analysis is performed for the recent observation of $\zcs$ exotic state in the $\ee$ annihilation experiment with motivation for measuring its spin quantum number in the future. Starting with the unpolarized electron and positron beam,
the polarization transfer to the $\zcs$ state and its decay angular distribution patterns are investigated. Some observables are suggested for determination of the spin parity quantum numbers. An ensemble of Monte-Carlo events are used to show some moment distributions special for manifestation of the different $\zcs$ spin scenarios.
\end{abstract}
\maketitle
\section{Introduction}
Search for exotic hadronic states, such as tetraquark or pentaquark states, receives much more attention both in experimental and theoretical physics community in recent years\cite{ali,olsen,yuancz}. Some exciting observation in experiment have been reported from the $\ee$ collider or hadron collider experiments. Most impressive among these are the charged or neutral $Z_c(3900)$ and $Z_c(4020)$ states, reported by the BESIII, Belle and CLEO collaborations \cite{zc3900,zc4020,beszc2,beszc3,beszc4,beszc5,beszc6,beszc7,beszc8,beszc9}. These observations open a new era for study of exotic hadronic spectroscopy.

Very recently, a new structure, dubbed $Z_{cs}(3985)^-$, is observed in the threshold enhancement of $D_s^-D^{*0}$ or $D_s^{*-}D^{0}$ invariant mass with significance of 5.3 standard deviation\cite{beszcs}. The pole position is determined to be $(3982.5^{+1.8}_{-2.6}\pm2.1)-{i\over 2}(12.8^{+5.3}_{-4.4}\pm3.0)$ MeV. This structure is suggested as a candidate of $c\bar cs\bar u$ tetraquark state, as predicted by the theoretical models \cite{theozcs}.

The spin and parity of $Z_{cs}(3985)^-$ is not determined experimentally due to limited events observed in the BESIII data. It was conjured as a $J^P=1^+$ state in the experimental determination of its detection efficiency. Theoretically, it was investigated as $J^P=0^+,1^+$ or $2^+$ state in the molecular and tetraquark state scenarios \cite{wangqn}. In this work, we perform the polarization analysis with a motivation to suggest some spin observables for the spin and parity measurement for this structure.

\section{Polarization analysis}
We consider an exact experiment of $\zcs$ production as observed in the unpolarized $\ee$ collisions. The center-of-mass (CM) energy of $\ee$ beams is set at about $\sqrt s=4.7$ GeV, so the $\zcs$ production is dominated by the electric magnetic process, from which the $\ee$ annihilate into a virtual photon $(\gamma^*)$, then it couples the $K^+\zcs^-$ final states. Without loss of generality, we assume that we detect the $\zcs$ state with a decay mode $\zcs^-\to D_s^-D^{*0}$.

\begin{figure}[htbp]
\centering
\includegraphics [width=8cm]{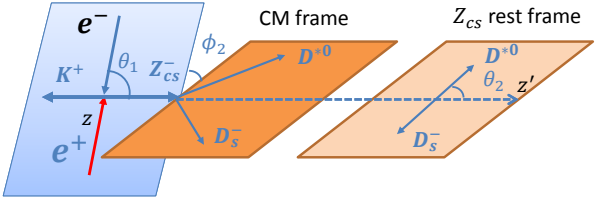}
\caption{(Color online) Orientation of helicity axes describing the $\zcs$ production and decay.  \label{sysdef}
}
\end{figure}

It is convenient for us to perform the $\zcs$ spin analysis and investigate the polarization transfer using the method of helicity amplitude. The $\zcs$ production and decay are described with helicity angles and amplitude as defined in Table \ref{tab::angdef}. The angles $\theta_1$ and $\phi_1$ are defined as the polar and azimuthal angles of $\zcs$ momentum in the $\ee$ CM system, with $z$ axis taken along the positron moving direction. While the angle $\theta_2$ is spanned by the three momentum of $\zcs$ and $D^{*0}$, here $D^{*0}$ is boosted to the $\zcs$ rest frame, and $\phi_2$ is the angle spanned by the $\zcs$ production and decay planes, as shown in Fig. \ref{sysdef}.

\begin{table}[htbp]
\caption{Definition of helicity angles and amplitudes for the $\zcs$ production and decay, here $\lambda_i(i=1,2)$ denotes the helicity values of corresponding particle.} \label{tab::angdef}
\centering{
\begin{tabular}{ccc}
\hline\hline
Decay & Angles & Amplitudes \\\hline
$\ee\to\gamma^*\to K^+\zcs^-(\lambda_1)$ & $\Omega_1(\theta_1,\phi_1)$ & $A_{\lambda_1}$\\
$\zcs^-\to D_s^- D^{*0}(\lambda_2)$ & $\Omega_2(\theta_2,\phi_2)$ & $B_{\lambda_2}$\\
\hline\hline
\end{tabular}
}
\end{table}

We perform the $\zcs$ spin analysis by calculating its spin density matrix (SDM), which encodes the complete polarization information and allow us to access it by studying the angular distribution of its final states $D^{*0}D_s^-$. Starting with the process of $\ee$ annihilation into a virtual photon, one has the $\gamma^*$ SDM as $\rho(\gamma^*)=$diag\{1,0,1\}, since this electric magnetic process conserves the parity and helicity. One can see that the $\gamma^*$ SDM deviates from the unit matrix, which indicates that the virtual photon has some degree of tensor polarization, while it is not longitudinally polarized.

The polarization transfer to the $\zcs$ can be estimated by calculation of the SDM in a straightforward manner with the method of helicity amplitude, under an assumption that $\zcs$ has the spin and parity $J^P=0^-,1^\pm,2^\pm$ or $3^\pm$. To note that the $0^+$ assignment is not allowed due to the fact that it violates the spin and parity in the process $\ee\to\gamma^*\to K^+\zcs^-$. In the space of $\zcs$ spin configuration, calculation of $\zcs$ SDM, $\rho(\zcs)$, can be visualized as the linear transformation with the decay matrix $M$. It reads \cite{ohlsen}
\begin{equation}\label{}
\rho(\zcs) = M\cdot \rho(\gamma^*)\cdot M^\dag,
\end{equation}
with $M_{\lambda_1,m}=D^1_{m,\lambda_1}(\phi_1,\theta_1,0)A_{\lambda_1}$, here $D^J_{m,\lambda_1}$ is an element of Wigner $D$-matrix. The number of independent helicity amplitude $A_{\lambda_1}$ is constrained by the helicity conservation law. It yields the relations
\begin{equation}\label{}
A_{-\lambda_1}=P(-1)^{J-1}A_{\lambda_1},
\end{equation}
for the $J^P$ assignment to the $\zcs$ state. Due to the constraint imposed by the $D^1_{m,\lambda_1}$ matrix, the helicity amplitudes will be vanishing if $\lambda_1>1$. This means that we are only left with two nonvanishing components $A_0$ and $A_1$. For the assignments of $1^-,2^+$ and $3^-$, the above parity conserving relation will further yield $A_0=0$.

Information on a particle polarization is often characterized with a set of multipole parameters $t^L_M$, which are used to define its SDM in terms of the spherical tensor operators $T^L_M$. Due to the property of non-Hermitian $T^L_M$ and the complex numbers $t^L_M$, it is not closely to relate the $t^L_M$ parameters to the experimental measurements. Here we choose an alterative Hermitian bases, $Q^L_M$, and real mulitipole parameters $r^L_M$ to form the SDM for a spin-$J$ particle \cite{qmatrix}. It reads

\begin{equation}\label{Qmat}
\rho (\zcs)=\dfrac {r ^{0}_{0}}{2J+1} \left( I+2J\sum ^{2J}_{L=1}\sum ^{L}_{M=-L}r ^{L}_{M}Q^{L}_{M}\right),
\end{equation}
where $I$ is a unit matrix with $(2J+1)\times(2J+1)$ dimensions, and $r^0_0$ is a unpolarization cross section, with $r^0_0=\text{Tr}\rho(\zcs)$.

To form a SDM for the spin-$J$ particle, one needs at the most the $J(J+2)$ real parameters, $r^L_M$, with the highest rank $L=2J$. The number of independent parameters $r^L_M$ will be greatly reduced if one imposes the parity conservation to the $\zcs$ production process and the rank condition to the calculation of $\zcs$ SDM. The parity conservation yields the symmetry relation
\begin{equation}\label{sdmsym}
\rho_{-m,-m'}=(-1)^{m-m'}\rho_{m,m'}.
\end{equation}
The rank condition requires that the rank of $\zcs$ SDM should be less or equal to that of virtual photon, namely
$$
\text{Rank}(\rho(\zcs))\le 2.
$$
The lists of nonvashing real parameters $r^L_M$ are given in Table \ref{rlist} and their expressions in terms of helicity amplitudes are given in appendix \ref{app:0m}-\ref{app::3m}.

%%%%%%%% table
\begin{table}[htbp]
\centering{
\caption{List of real multipole parameters ($r^L_M$) and analyzing powers ($A^L_M$) for different $J^P$ assignments to the $\zcs$ particle.}\label{rlist}
\begin{tabular}{c|cccccc}
\hline\hline
~~$J^P$~~ &~~ $1^+$ ~~&~~ $1^-$ ~~&~~ $2^+$ ~~&~~$2^-$~~&~~$3^+$ ~~&~~$3^-$\\\hline
 &$r^2_{0,1,2}$&$r^2_{0,2}$&$r^2_{0,2}$&$r^2_{0,1,2}$&$r^2_{0,1,2}$&$r^2_{0,2}$ \\
$r^L_M$&&&$r^4_{0,2}$&$r^4_{0,1,2}$&$r^4_{0,1,2}$&$r^4_{0,2}$\\
&&&&&$r^6_{0,1,2}$&$r^6_{0,2}$\\\hline
 &$\mathcal A^2_{0,1,2}$&$\mathcal A^2_{0,2}$&$\mathcal A^2_{0,2}$&$\mathcal A^2_{0,1,2}$&$\mathcal A^2_{0,1,2}$&$\mathcal A^2_{0,2}$ \\
$\mathcal A^L_M$&&&$\mathcal A^4_{0,2}$&$\mathcal A^4_{0,1,2}$&$\mathcal A^4_{0,1,2}$&$\mathcal A^4_{0,2}$\\
&&&&&$\mathcal A^6_{0,1,2}$&$\mathcal A^6_{0,2}$\\
\hline\hline
\end{tabular}
}
\end{table}
%%%%%%%%

The degree of $\zcs$ polarization measures the departure of its SDM from isotropy. We take the definition for the spin-$J$ particle as \cite{eleader}
\begin{equation}\label{}
d={1\over \sqrt{2J}}\left[(2J+1)\text{Tr}\tilde\rho^2-1\right]^{1/2},
\end{equation}
where $\tilde\rho$ is the normalized SDM, and its angular dependence has been integrated out. With the obtained SDM, the degree of polarization for different $J^P$ assignment to $\zcs$ are calculated to be
\begin{equation}
d=\left\{\begin{array}{lc}
   {\sqrt7\over4}& \text{ for $1^-$}, \\
   {1\over 2}\sqrt{4+r(7r-8)\over (1+2r)^2}& \text{ for $1^+$}, \\
   {1\over 4}\sqrt{17\over 2}& \text{ for $2^+$}, \\
   {1\over 2\sqrt2}\sqrt{8-8r+17r^2\over (1+2r)^2}& \text{ for $2^-$}, \\
   {1\over 2\sqrt3}\sqrt{12+r(27r-8)\over (1+2r)^2}& \text{ for $3^+$}, \\
   {3\over4}& \text{ for $3^-$},
\end{array}\right.
\end{equation}
where $r=|A_1/A_0|$. One can see that the $\zcs$ has a degree of polarization larger than 60\% for the assignment of $J^P=1^-,2^+$ and $3^-$. The $\zcs$ polarization originates from the polarization transfer of the transverse polarization of virtual photon.

To access the $\zcs$ polarization information, one needs to study the implication of the $D^{*0}$ angular distribution in its decaying final states. Thus the decay $\zcs^-\to D_s^-D^{*0}$ serves as a polarimetry. Likewise we investigate the polarization transfer from the $\zcs$ to $D^{*0}$ by calculation of the SDM for the $D^{*0}$ particle. It reads

\begin{equation}\label{}
\rho(\dst)=N\cdot\rho(\zcs)\cdot N^\dag,
\end{equation}
with decay matrix $N_{\lambda_2,\lambda_1}=B_{\lambda_2}D^J_{\lambda_1,\lambda_2}(\phi_2,\theta_2,0)$.
Then the joint angular distribution for the two sequential decay can be calculated by taking trace of $\rho(\dst)$, thus one has
\begin{equation}\label{jad}
\mathcal{I}(\Omega_1,\Omega_2)=\mathcal{I}_0\left(1+\sum_{L=1}^{2J}\sum_{M=-L}^{L}r^L_M \mathcal A^L_M\right)
\end{equation}
with
\begin{eqnarray}
\mathcal{I}_0&=&{r^0_0\over 2J+1}\text{Tr}N_fN_f^\dag,\\
\mathcal A^L_M&=&{2J\text{Tr}N_fQ^L_MN_f^\dag\over \text{Tr}N_fN_f^\dag},\\
\mathcal A^0_0&=&\text{Tr}N_fN_f^\dag,
\end{eqnarray}
where $\mathcal{I}_0$ is the cross section corresponding to the unpolarization case, and $\mathcal A^L_M$ is always known as the analysing power. The list of nonvanishing $\mathcal A^L_M$ is given in Table \ref{rlist}, and their expressions in terms of amplitudes $B_{\lambda_2}$ and angles $\Omega_2$
are given in appendix \ref{app:0m}-\ref{app::3m}.

To figure out the $\zcs$ spin and parity, an intuitive way is to check the joint angular distribution in respect of $\cos\theta_1$ and $\cos\theta_2$. It is easy to get it by integrating out the $\phi_2$ in the $\mathcal I(\Omega_1,\Omega_2)$ distribution. Then one has

\begin{figure}[htbp]
\centering
\includegraphics [width=8cm]{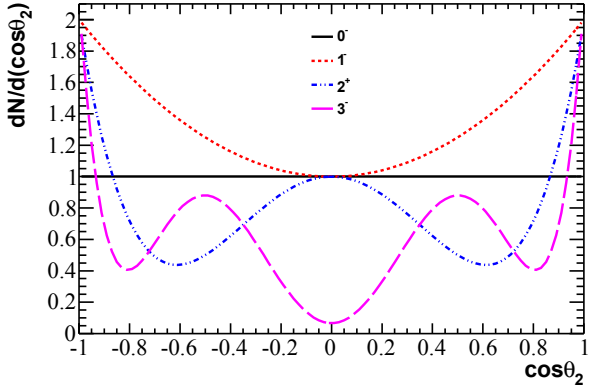}
\caption{(Color online) Angular distribution of ${dN\over d\cos\theta_2}$ for the spin and parity assignments $J^P=1^-,2^+$ and $3^-$. \label{cos2}
}
\end{figure}

\begin{widetext}
\begin{equation}\label{cos12}
{dN\over dx_1dx_2}\propto \left\{
\begin{array}{lc}
(3+5 r_2)+(3+r_2) x_1^2+(r_2-1) (5 x_1^2+1)x_2^2,&\text{ for $1^+$}\\
(1+x_1^2)(1+x_2^2),&\text{ for $1^-$}\\
(1+x_1^2) (1-3x_2^2+4 x_2^4),&\text{ for $2^+$}\\
2 r_1 \left(x_1^2+1\right) \left[\left(4 r_2-3\right) x_2^4-3 \left(r_2-1\right)
   x_2^2+r_2\right]&\nonumber\\
   +\left(x_1^2-1\right) \left[3 \left(4 r_2-3\right) x_2^4+\left(6-12 r_2\right)
   x_2^2-1\right],&\text{ for $2^-$}\\
 4 \left(x_1^2-1\right) [3 r_2 \left(1-5 x_2^2\right){}^2 \left(x_2^2-1\right)-2 \left(5 x_2^3-3 x_2\right){}^2]&\nonumber\\
+r_1\left(x_1^2+1\right) [r_2 \left(225 x_2^6-305 x_2^4+111 x_2^2+1\right)-6 \left(1-5 x_2^2\right){}^2 \left(x_2^2-1\right)],&\text{ for $3^+$}\\
 \left(1+x_1^2\right) \left(1+111 x_2^2-305 x_2^4+225 x_2^6\right),&\text{ for $3^-$}
\end{array}\right.
\end{equation}
where $x_1=\cos\theta_1,~x_2=\cos\theta_2,~r_1=|A_1/A_0|^2,~r_2=|B_1/B_0|^2$.
\end{widetext}
For the cases of $1^-,~2^+$ and $3^-$ assignments to the $\zcs$ state, one can see that the angular distributions are factored out from the undetermined helicity amplitudes in the above equations, and their patterns are ambiguously determined according to the $\zcs$ spin and parity quantum numbers. Due to the factor that the virtual photon is produced from the $\ee$ annihilation, and the vector coupling to the $e^-/e^+$ Dirac spinor conserves their helicities, the component of helicity zero is negligible. This leads to deviation of the spin density matrix of virtual photon from the unit matrix. This is equivalent to the description of tensor polarization for the vector particles, which is further manifested in the nontrival pattern of the $\zcs$ angular distributions. Especially, the polarization transfer to the $\zcs$ state leads to that it acquires some degree of longitudinal polarizations, and this gives rise to the unflat profile of angular distribution for the decayed particles. In experiment, one can get knowledge about these quantum numbers by interpreting the implication of angular distribution ${dN\over dx_2}$, which was shown in Fig. \ref{cos2} for the $J^P=1^-,~2^+$ and $3^-$ cases.

For the cases of $J^P=1^+,~2^-$ and $3^+$ assignments, the angular distribution is dependent on the amplitudes ratios of $r_1$ and $r_2$, which are related to the study of polarization transfer to the $\zcs$ state. However, the polarization information is inaccessible directly in the modern electromagnetic spectrometers, since they are often designed with a general purpose to detect the charged particle and neutral showers for high energy experiments. In the considered $\ee$ collision experiment, we express the spin transfer to the $\zcs$ states with a set of real multipole parameters, $r^L_M$, as we list in Table \ref{rlist}. Nonetheless, one can study the implication of analysing power for the subsequent decay, $\zcs\to D_sD^{*0}$, to get the polarization information. In experiments, one can require the mass of $D_sD^{*0}$ falling into the $\zcs$ mass region, and looking at the distribution of an analysing power in respect of the $\zcs$ polar angle. This is equivalent to looking at the distribution of first moments $\langle  a^L_M \rangle \sim \cos\theta_1$, here $a^L_M$ is the angular dependent factor of the analysing power $\mathcal A^L_M$.

Using the joint angular distribution $\mathcal I(\Omega_1,\Omega_2)$, we get
\begin{equation}\label{}
\langle a^L_M \rangle \sim \int a^L_M(\Omega_2)\mathcal I(\Omega_1,\Omega_2)d\Omega_2\sim r^L_M(x),
\end{equation}
with $x=\cos\theta_1$. Explicitly, we list the first moments distribution $\langle a^L_M\rangle$ versus $x$ in Table \ref{table::analist}. For the components $M=0$, one has the first moments $\langle ~a^L_0\rangle=1+\alpha_L x$ with $L=2,4,6$, here $\alpha_L$ is an angular distribution parameter, determined by the helicity amplitudes $A_{\lambda_1}$. They are given in the appendix \ref{app::angparlist}.

\begin{widetext}
\begin{table*}[htbp]
\centering{
\caption{The analysing power $a^L_M$ and its first moments distribution $\langle a^L_M\rangle$ in terms of $x=\cos\theta_1$. The symbols $\surd$ below the $J^P$ assignment indicate that it has the corresponding analysing power. }\label{table::analist}
\begin{tabular}{l|c|cccccc}
 \hline \hline
  $a^L_M$ &~~$\langle a^L_M\rangle $~~& ~~$J^P=1^+$~~ & ~~$1^-$~~ & ~~$2^+$~~ & ~~$2^-$~~& ~~$3^+$~~ & ~~$3^-$~~\\\hline
  $a^2_2= \sin ^2\theta _2 \cos \left(2 \phi_2\right)$&$1-x^2$&$\surd$&$\surd$&$\surd$&$\surd$&$\surd$&$\surd$ \\
  $a^2_1=\sin \left(2 \theta _2\right) \cos \left(\phi _2\right)$&$x\sqrt{1-x^2}$&$\surd$&&&$\surd$&$\surd$& \\
  $a^2_0=3\cos \left(2 \theta _2\right)$&$1+\alpha_2 x^2$&$\surd$&$\surd$&$\surd$&$\surd$&$\surd$&$\surd$ \\\hline
  $a^4_2=\sin ^2\theta _2 \left[7 \cos \left(2 \theta _2\right)+5\right]
   \cos \left(2 \phi _2\right)$&$1-x^2$&&&$\surd$&$\surd$&$\surd$&$\surd$ \\
  $a^4_1= \sin \theta _2\left[9 \cos \left(\theta _2\right)+7 \cos \left(3
   \theta _2\right)\right] \cos\phi _2$&$x\sqrt{1-x^2}$&&&&$\surd$&$\surd$& \\
  $a^4_0=20 \cos \left(2
   \theta _2\right)+35 \cos \left(4 \theta _2\right)$&$1+\alpha_4x^2$&&&$\surd$&$\surd$&$\surd$&$\surd$ \\\hline
  $a^6_2=\sin^2\theta _2  \left[60 \cos \left(2 \theta
   _2\right)+33 \cos \left(4 \theta _2\right)+35\right]\cos \left(2 \phi
   _2\right)$&$1-x^2$&&&&&$\surd$&$\surd$ \\
  $a^6_1=\left[5 \sin \left(2 \theta _2\right)+12 \sin \left(4 \theta
   _2\right)+33 \sin \left(6 \theta _2\right)\right] \cos \left(\phi _2\right)$&$x\sqrt{1-x^2}$&&&&&$\surd$& \\
  $a^6_0=105 \cos \left(2
   \theta _2\right)+126 \cos \left(4 \theta _2\right)+231 \cos \left(6 \theta
   _2\right)$&$1+\alpha_6x^2$&&&&&$\surd$&$\surd$ \\
  \hline\hline
\end{tabular}
}
\end{table*}
\end{widetext}
\section{Applications}
We consider the discovery process of $\ee\to K^+\zcs^-,~\zcs^-\to D_s^-D^{*0}$, in which the $\zcs(3985)^-$ state was observed for the first time in the recoiling mass spectrum of $RM(K^+)$ \cite{beszcs}. The mass and width of the $\zcs$ state was measured based on the 85 observed signal events. Analysis on the angular distributions needs a large size of data events. In present, the hints on the spin and parity numbers can only be investigated from the mass spectrum based on the amplitude model, using the dependence of the lineshape of $\zcs$ state on the spin and parity numbers.
\subsection{Amplitude model}
We construct the amplitude model with the method of the covariant tensor formalism, in which we consider the hypotheses the $\zcs$ state has the spin and parity numbers as $J^P=0^-,~1^\pm,~2^\pm$ and $3^\pm$, respectively. The possibility of other higher spin assignments should be depressed by the factor of the centrifugal barrier. We take the $0^-$ hypothesis as an example to illustrate the construction of amplitude model. For the first decay $\gamma^*(p_1)\to \zcs^-(p_2)K^+(p_3)$, here $p_1,~p_2$ and $p_3$ denote the momenta of the corresponding particle,
only one partial wave, $L=1$, is allowed due to the decay conserves the spin and parity. The form of coupling vertex is taken as $f_2p_3^\mu$ with a complex coupling constant $f_2$. For the second decay $\zcs^-(p_2)\to D_s^-(p_4)D^{*0}(p_5)$, the coupling vertex can be written as $f_5p_5^\nu$ with a complex parameter $f_5$. Then the joint amplitude reads
\begin{equation}
\mathcal{M}_{0^-}={f_2f_5(\epsilon^*\cdot p_3)(\epsilon_{D^*}\cdot p_5)\over m_{D_sD^{*}}^2-M^2-iM\Gamma},
\end{equation}
where $\epsilon(\epsilon_{D^*})$ is the polarization vector for the virtual photon ($D^{*0}$).

The tensor forms of coupling vertexes for other hypotheses, $J^P=1^\pm,2^\pm$ and $3^\pm$, are summarized in Table \ref{Tab::vertex}, which are constructed with the Levi-Civita tensor $\epsilon_{\alpha\beta\mu\nu}$ and $p^\mu$, respecting the $CP$ conservation. The joint amplitude for the two-step sequential decay is constructed by multiplying the propagator to contract the Lorentz indexes of the $\zcs$ resonance. The propagators are parametrized with function of relativistic Breit-Wigner, and they are given in Appendix I. It is worthy noting that the structure of tensor $\tilde{g}_{\mu\nu}$ in the Breit-Wigner is dependent on the spin assignment to the $\zcs$ state, this leads to the lineshape of $\zcs$ state dependent on the spin quantum numbers. This is the idea we adopt to investigate the spin hints from the mass spectrum, but the conclusive spin and polarization analysis is needed to directly perform the angular analysis if the large size of data events available in the further.

One feature in the joint amplitude is that the coupling constants, $f_i$, can be factorted out from the part of tensor amplitude for hypotheses $J^P=0^-,~1^-,~2^+$ and $3^-$. This is equivalent to the separation of the angular distribution as discussed in the polarization analysis with the helicity amplitude. This leads to the fact that the spin analysis for these four assignments is model independent. However, to distinguish them from other spin assignments, ie. $J=1^+,~2^-$ and $3^+$, one needs to do amplitude analysis to determine the coupling constants by fitting the mass spectrum of data events.

%%%%%%%%%%% table %%%%%%%%%%%
\begin{widetext}
\begin{table*}[htbp]
\begin{center}
\caption{The tensor forms of coupling vertexes for the two decays. Here $p_i(i=1,...,5)$ are the momenta of corresponding particles. The symbols $f_i$ denote the coupling constants, and they are taken as complex numbers. The Greek letters are the Lorentz indexes.}
\label{Tab::vertex}
\begin{tabular}{ccc}
\hline\hline
Decay~~~~~~~ &  $\gamma^*(p_1)\to \zcs^-(p_2)K^+(p_3)$ &$\zcs^-(p_2)\to D_s^-(p_4)D^{*0}(p_5)$\\\hline
$0^-$ &  $f_2p_3^\mu$ & $f_5p_5^\nu$\\\hline
$1^-$ &  $-f_{10}\epsilon_{\alpha\beta\mu\nu}p_2^\alpha p_3^\beta$ & $-f_{11}\epsilon_{\alpha\beta\mu'\nu'}p_2^\alpha p_4^\beta$\\\hline
$1^+$ & $p_{3\mu}p_{3\nu}f_6+g_{\mu\nu}f_7$ &  $p_{4\mu'}p_{4\nu'}f_8+g_{\mu'\nu'}f_9$\\\hline
$2^-$ & $p_{3\mu}p_{3\nu}p_{3\alpha}f_{18}+p_{3\mu}g_{\nu\alpha}f_{19}$&$p_{4\mu'}p_{4\nu'}p_{4\alpha'}f_{20}+p_{4\mu'}g_{\nu'\alpha'}f_{21}$\\\hline
$2^+$ & $-\epsilon_{\lambda\beta\mu\alpha}p_{1\lambda}p_{2\beta}p_{3\nu}f_{16}$&$-\epsilon_{\lambda\beta\mu'\alpha'}p_{2\lambda}p_{4\beta}p_{4\nu'}f_{17}$\\\hline
$3^-$ &$-\epsilon_{\lambda\eta\mu\beta}p_{1\lambda}p_{3\eta}p_{3\nu}p_{3\alpha}f_{28}$&$-\epsilon_{\lambda\eta\mu'\beta'}p_{2\lambda}p_{4\eta}p_{4\nu'}p_{4\alpha'}f_{29}$\\\hline
$3^+$ & $p_{3\mu}p_{3\nu}p_{3\alpha}p_{3\beta}f_{24}+p_{3\mu}p_{3\nu}g_{\alpha\beta}f_{25}$& ~~~~~~~~~~$p_{4\mu'}p_{4\nu'}p_{4\alpha'}p_{4\beta'}f_{26}+p_{4\mu'}p_{4\nu'}g_{\alpha'\beta'}f_{27}$\\\hline\hline

\end{tabular}
\end{center}
\end{table*}
\vspace{0.5cm}
\end{widetext}
%%%%%%%%%%%%%%%%%%%%%%%%%%%%

\subsection{Fit to data events}
We determine the coupling constant for the spin assignments, $1^+,~2^-$ and $3^+$, using the mass spectrum as reported by the BESIII collaboration \cite{beszcs}. The $\zcs$ state was searched in the four data sets within the $\ee$ center-of-mass energy from $\sqrt s=4.681\sim4.698$ GeV. Only the significant $\zcs$ state was observed at the first data set. We take the recoiling mass spectrum of $K^+$ at this energy point to do the amplitude analysis. The combinatory background shape is modeled by the probability density distribution
\begin{equation}
M^\text{bg}_i=g_\text{bg}\exp[-a\cdot m_i-b\cdot m_i^2],
\end{equation}
where $m_i$ is the observed recoil mass of $K^+$ in the $i$-th bin, and the parameters $g_\text{bg}=1.07\pm0.11~\text{GeV}^{-1/2},~a=-6.44\pm1.08~\text{GeV}^{-1}$ and $b=1.37\pm0.34~\text{GeV}^{-2}$ are determined by fitting the lineshape of combinatory background events as reported in the experiment. Then the observed events in the $i$-th bin can be calculated with
\begin{eqnarray}\label{sigyield}
N_i^{\mathrm{th}} &=&c_0( M_i^{\mathrm{bg}}+\int \frac{1}{(2\pi)^5}\frac{1}{16s} \overline{|\mathcal{M}_{J^P}|}^2 |\vec{p}^\ast_{D_s^-}| |\vec{p}_{K^+}|\nonumber\\
&\times& d\Omega^\ast_{D_s^-} d\Omega_{K^+}),
\end{eqnarray}
where $c_0$ is an overall parameter, $\mathcal{M}_{J^P}$ is the amplitude corresponding to the spin and parity assignments to the $\zcs$ state. With $s$ being the center of mass energy, $|\vec{p}^\ast_{D_s^-}|$ and $\Omega^\ast_{D_s^-}$ are the momentum and solid angle of $D_s^-$ in the rest frame of $D^\ast D_s^-$ system, respectively. $|\vec{p}_{K^+}|$ and $\Omega_{K^+}$ are the momentum and solid angle of $K^+$ respectively, and the overline means the spin average over the initial $\gamma^\ast$ and the spin sum over the final $D^\ast$.

We determine the coupling constants, $f_i$, using the least square method. The object function to be minimized is defined as
\begin{equation}
\chi^2 = \sum_{i} {(N^\text{dt}_i - N^\text{th}_i)^2\over N^\text{dt}_i},
\end{equation}
where $N^\text{dt}_i$ is the number of observed events in the $i$-bin, and the uncertainty of observed events follows the Poisson distribution as reported in experiment, and the sum runs over all bins in the fit range.

Since the $\zcs$ state was observed at the mass threshold of $D_s^-D^{*0}/D^{*-}D^{0}$, and the reported lineshape above 4.02 GeV can be explained with the combinatory background events, we limit the fit range to $RM(K^+)<4.02$ GeV. We determine the coupling constants $f_i$ for the each spin assignment $1^+,~2^-$ or $3^+$, in which we have 4 parameters in the joint amplitude. Due to that the number of observed events is defined to be proportional to the density of amplitude, it will introduce a nonphysical parameter in the Eq. (\ref{sigyield}). This implies that only the relative values of coupling constants can be chosen as the physics parameters in the fit. From Table \ref{Tab::vertex}, we chose the parameters $f_6,~f_8,~f_{18},~f_{20},f_{24}$ and $f_{26}$ as reference parameters by fixing the value at 1 in the fit. In this way, the other determined parameters are understood as the relative values to the reference parameters.

In the fit, the mass and width of the $\zcs$ state is fixed to the reported values for each spin and parity assignment, namely $M=3982.5$ MeV and $\Gamma=12.8$ MeV. The parameters are obtained by minimizing the $\chi^2$ object function, and they are determined to be $f_7\approx0.186\pm0.035,~f_9\approx-0.074\pm0.018$ for the hypothesis $J^P=1^+$, and $f_{19}=0.299\pm0.036,~f_{21}=-0.278\pm0.067$ for the $2^-$ case, and $f_{25} =0.012\pm0.002,~f_{27}=0.297\pm0.047$ for the $2^-$ case. Here we take the coupling parameters as real numbers. In general, they are able to take as complex numbers. But this choice will introduce some redundant parameters, relative to the limited data events available.

\begin{figure*}[htbp]
\centering
\includegraphics [width=16cm]{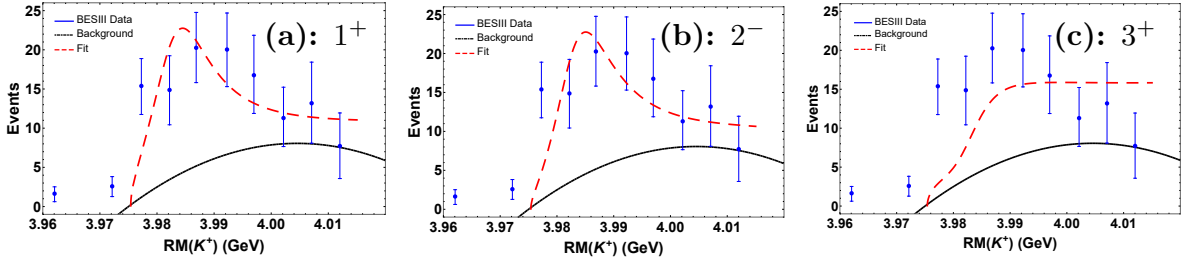}
\caption{(Color online) Fit results of the recoil mass $RM(K)$ for the different hypotheses of $\zcs$ spin and parity, (a) $1^+$, (b) $2^-$ and (c) $3^+$. The data events and background shape are quoted from Ref. \cite{beszcs}. \label{fig::fitJp}
}
\end{figure*}

The fit results are displayed in Fig. \ref{fig::fitJp}, and they are in good agreement with the data events for the $1^+$ and $2^-$ hypotheses. The fit goodness for these two cases are almost the same, but it gets worse for the $3^-$ hypothesis. The lineshape for the $3^-$ hypothesis gets flat above the $\zcs$ mass, since the momentum dependence in the Breit-Wigner function introduce more high power terms in the lineshape, and this leads to the enhancement with the mass increase. In view of fit goodness, the possibility of $3^-$ assignment can be ruled out.

\subsection{Monte-Carlo simulations}
\begin{figure}[htbp]
\centering
\includegraphics [width=8cm]{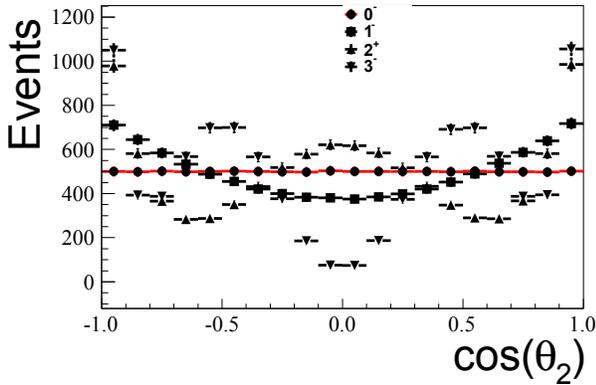}
\caption{(Color online) Angular distribution of generated events for the spin and parity assignments $J^P=0^-,1^-,2^+$ and $3^-$. \label{fig::cos2}
}
\end{figure}

We generate an ensemble of Monte-Carlo events using the amplitude model, with the coupling constant fixed to the determined values. Figure \ref{fig::cos2} shows the helicity angular distribution $\cos\theta_2$ for the assignments of $J^P=0^-,~1^-,~2^+$ and $3^-$. One can see that these distributions are distinguishable from each other, and they are consistent with that shown in Fig. \ref{cos2}. To compared with the pattern of observed angular distribution, one can figure out the spin and parity assignment of $\zcs$ state. The advantage using these distributions is that they are model independent, and are easily checked in experiment.

\begin{figure*}[htbp]
\centering
\includegraphics [width=17cm]{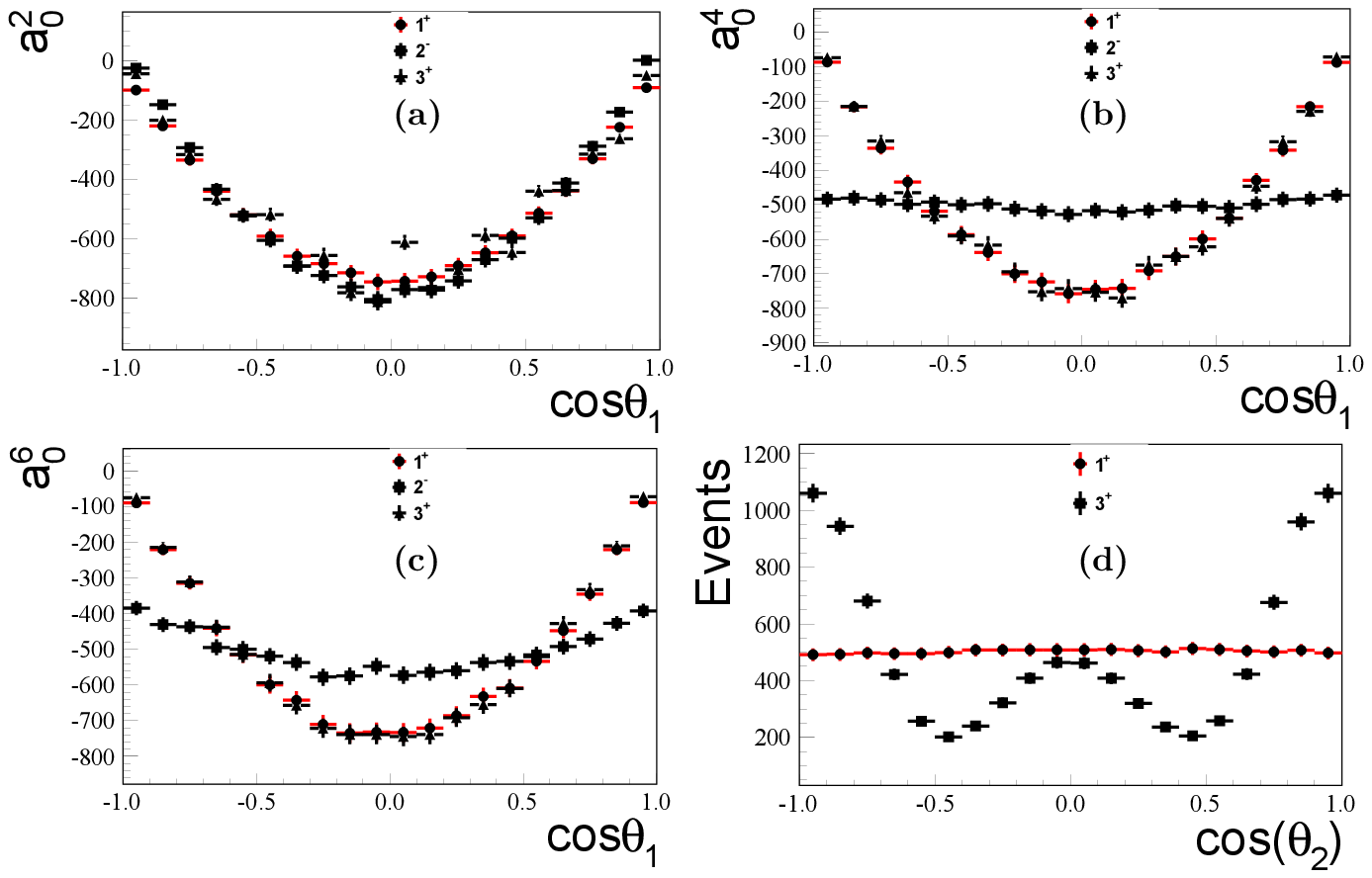}
\caption{(Color online) Distributions of the first moments $a^L_M$ versus $\cos\theta_1$ for the different spin and parity $(J^P)$ assignments. The plots are filled with a set of Monte-Carlo events as described in the text. \label{fig::moments}
}
\end{figure*}

However, to distinguish the hypotheses of spin and parity $1^+,~2^-$ and $3^+$, one needs an analysis of data events, especially the angular distribution. With the coupling constants obtained from the fit to the data spectrum, we present some distributions with Monte-Carlo events as shown in Fig. \ref{fig::moments}, serving as the footprint mark. If we check the moment $a_2^0$ distribution, the distributions are almost indistinguishable for these three $J^P$ assignments. One needs further checking the distribution of high moments, such as $a_4^0$ and $a_6^0$, and the $2^-$ case can be figured out. Then we can further distinguish the $1^+$ from $3^+$ hypothesis by checking the $\cos\theta_2$ distribution as shown Fig.  \ref{fig::moments}(d).

%%%%%%%%%%%%%%%%%%% figure
%%%%%%%%%%%%%%%%%%%%%%%%%%%
\section{Summary}
We carry out an analysis on the $\zcs$ polarization for the motivation of measuring the $\zcs$ spin and parity in the future experiment. We consider the production process $\ee\to\gamma^*\to K^\pm\zcs^\mp,~\zcs^\mp\to D_s^-D^{*0}$ or $D_s^{*-}D^{0}$ at the $\ee$ annihilation experiment. As the spin and parity conservation is concerned for the $\zcs$ production, the quantum numbers of $J^P=0^+$ are not allowed for the virtual photon coupled to the $K^\pm\zcs^\mp$ states. Hence the analysis is performed for the scenarios of $J^P=0^-,1^\pm$ and $2^\pm$.  Due to fact that the virtual photon from the $\ee$ annihilation is of tensor polarization, the electromagnetic coupling to the $\zcs$ state will transfer some degree of polarization to the $\zcs$ of nonzero spin state. The subsequential $\zcs$ decay can be used as polarimeter to manifest its spin and parity by studying the angular distribution of decayed particles. The spin assignments $0^-,~1^-,~2^+$ and $3^-$ can be distinguishable by checking the angular distribution of the $\zcs$ decayed particle, which are model independent. While to figure out the $1^+,~2^-$ and $3^+$ hypotheses, one needs performing the amplitude analysis. With a few data events available, we perform a fit to the mass spectrum and get hints for these spin hypotheses. Using obtained parameters in the amplitude model, an ensemble of toy MC events are generated, we present some distributions as footprint mark to figure out these spin hypotheses.  Some moments distributions are formulated, and suggested as polarization observables to manifest the $\zcs$ spin and parity in the experiment if a large size data sets is available in the future.

%\begin{widetext}
%\newpage
\acknowledgments
The work is partly supported by
the National Natural Science Foundation of China under Grants No. 11875226, 11875262 and 11835012.
~\\

%\newpage\newpage
\appendix

\section{$J^P=0^-$}\label{app:0m}
Multipole parameter $r_0^0$ is expressed in terms of helicity amplitude $A_0$ for the $\zcs$ $J^P=0^-$ hypothesis.
\begin{equation}
r_0^0=|A_0|^2 \sin^2\theta_1.\nonumber
\end{equation}

\section{$J^P=1^+$}
There are 4 multipole parameters ($r^L_M$) and analyzing powers ($\mathcal  A^L_M$) for the  $\zcs$ $J^P=1^+$ hypothesis. The definitions are referred to Table \ref{rlist}, and Eq. \ref{jad}. $A_{\lambda_1}$ and $B_{\lambda_2}$ are helicity amplitudes as defined in Table \ref{tab::angdef}.
\begin{eqnarray}
r^0_{0} &=&\frac{1}{4} \left\{2 |A_0|^2 \sin^2
   \theta_1+|A_1|^2\left[\cos \left(2 \theta
   _1\right)+3\right]\right\} \nonumber\\
r^0_0r^2_2 &=&\frac{\sqrt{3}}{4}  |A_1|^2 \sin^2
  \theta _1\nonumber\\
r^0_0r^2_1&=&-\frac{\sqrt{3}}{4}  \text{Re}(A_1A^*_0) \sin \left(2 \theta _1\right)\nonumber\\
r^0_0r^2_0&=&\frac{1}{8} \left\{|A_1|^2 \left[\cos
   \left(2 \theta _1\right)+3\right]-4 |A_0|^2 \sin^2\theta
   _1\right\}\nonumber.
\end{eqnarray}

\begin{eqnarray}
\mathcal  A^0_0&=&2|B_1|^2+|B_0|^2,\nonumber\\
\mathcal  A^0_0\mathcal  A^2_2&=&\sqrt{3} \left(|B_1|^2-|B_0|^2\right) \sin ^2\theta _2 \cos \left(2 \phi_2\right),\nonumber\\
\mathcal  A^0_0\mathcal  A^2_1&=&\sqrt{3} \left(|B_1|^2-|B_0|^2\right) \sin \left(2 \theta _2\right) \cos \left(\phi _2\right),\nonumber\\
\mathcal  A^0_0\mathcal  A^2_0&=&\frac{1}{2} \left(|B_1|^2 -|B_0|^2\right) \left[3 \cos \left(2 \theta _2\right)+1\right]\nonumber.
\end{eqnarray}
\section{$J^P=1^-$}
There are 3 multipole parameters ($r^L_M$) and analyzing powers ($\mathcal  A^L_M$) for the  $\zcs$ $J^P=1^-$ hypothesis. The definitions are referred to Table \ref{rlist}, and Eq. \ref{jad}. $A_{\lambda_1}$ and $B_{\lambda_2}$ are helicity amplitudes as defined in Table \ref{tab::angdef}.
\begin{eqnarray}
r^0_0&=&\frac{1}{4} |A_1|^2 \left[\cos \left(2 \theta _1\right)+3\right],\nonumber\\
r^0_0r^2_2&=&-\frac{1}{4} \sqrt{3} |A_1 |^2 \sin ^2\theta _1,\nonumber\\
r^0_0r^2_0&=&\frac{1}{8} |A_1|^2 \left[\cos \left(2 \theta _1\right)+3\right]\nonumber,
\end{eqnarray}
\begin{eqnarray}
\mathcal  A^0_0&=&2|B_1|^2,\nonumber\\
\mathcal  A^0_0\mathcal  A^2_2&=&\sqrt{3} |B_1|^2\sin ^2 \theta _2 \cos \left(2 \phi
   _2\right),\nonumber\\
\mathcal  A^0_0\mathcal  A^2_0&=&\frac{1}{2}|B_1|^2 \left[3 \cos \left(2 \theta _2\right)+1\right]\nonumber.
\end{eqnarray}

\section{$J^P=2^+$}
There are 5 multipole parameters ($r^L_M$) and analyzing powers ($\mathcal  A^L_M$) for the  $\zcs$ $J^P=2^+$ hypothesis. The definitions are referred to Table \ref{rlist}, and Eq. \ref{jad}. $A_{\lambda_1}$ and $B_{\lambda_2}$ are helicity amplitudes as defined in Table \ref{tab::angdef}.
\begin{eqnarray}
r^0_0&=&\frac{1}{4} |A_1|^2 \left[\cos \left(2 \theta _1\right)+3\right],\nonumber\\
r^0_0r^4_2&=&\frac{1}{2} \sqrt{\frac{5}{14}} |A_1|^2 \sin^2\theta _1,\nonumber\\
r^0_0r^4_0&=&-\frac{1}{2 \sqrt{14}}|A_1|^2 \left[\cos \left(2 \theta _1\right)+3\right],\nonumber\\
r^0_0r^2_2&=&-\frac{1}{4} \sqrt{\frac{15}{14}}| A_1|^2 \sin^2\theta _1,\nonumber\\
r^0_0r^2_0&=&-\frac{1}{8} \sqrt{\frac{5}{14}} | A_1|^2 \left[\cos \left(2 \theta
   _1\right)+3\right]\nonumber.
\end{eqnarray}

\begin{eqnarray}
\mathcal  A^0_0&=&2|B_1|^2,\nonumber\\
\mathcal  A^0_0\mathcal  A^4_2&=&-\sqrt{\frac{10}{7}} |B_1|^2 \sin^2\theta _2\left[7 \cos
   \left(2 \theta _2\right)+5\right]\cos \left(2 \phi _2\right),\nonumber\\
\mathcal  A^0_0\mathcal  A^4_0&=&-\frac{1}{4 \sqrt{14}}|B_1|^2 \left[20 \cos \left(2 \theta _2\right)+35 \cos \left(4
   \theta _2\right)+9\right],\nonumber\\
\mathcal  A^0_0\mathcal  A^2_2&=&-\sqrt{\frac{30}{7}} |B_1|^2 \sin ^2\theta _2 \cos \left(2
   \phi _2\right),\nonumber\\
\mathcal  A^0_0\mathcal  A^2_0&=&-\sqrt{\frac{5}{14}} |B_1|^2 \left[3 \cos \left(2 \theta
   _2\right)+1\right]\nonumber.
\end{eqnarray}
\section{$J^P=2^-$}
There are 7 multipole parameters ($r^L_M$) and analyzing powers ($\mathcal  A^L_M$) for the  $\zcs$ $J^P=2^-$ hypothesis. The definitions are referred to Table \ref{rlist}, and Eq. \ref{jad}. $A_{\lambda_1}$ and $B_{\lambda_2}$ are helicity amplitudes as defined in Table \ref{tab::angdef}.
\begin{eqnarray}
r^0_0&=&\frac{1}{4} \left\{2 |A_0|^2 \sin ^2\theta _1+|A_1|^2 \left[\cos \left(2 \theta _1\right)+3\right]\right\},\nonumber\\
r^0_0r^4_2&=&-\frac{1}{2} \sqrt{\frac{5}{14}}|A_1|^2 \sin ^2\theta _1,\nonumber\\
r^0_0r^4_1&=&\frac{1}{4} \sqrt{\frac{15}{7}} \text{Re}(A_1 A^*_0) \sin
   \left(2 \theta _1\right),\nonumber
\end{eqnarray}
\begin{eqnarray}
r^0_0r^4_0&=&\frac{3 |A_0|^2 \sin^2\theta _1-|A_1|^2
   \left[\cos \left(2 \theta _1\right)+3\right]}{2 \sqrt{14}},\nonumber\\
r^0_0r^2_2&=&\frac{1}{4} \sqrt{\frac{15}{14}} |A_1|^2\sin ^2\theta _1,\nonumber\\
r^0_0r^2_1&=&-\frac{1}{4} \sqrt{\frac{5}{14}} \text{Re}(A_1 A^*_0)
   \sin \left(2 \theta _1\right),\nonumber
\end{eqnarray}
\begin{eqnarray}
r^0_0r^2_0&=&-\frac{1}{8} \sqrt{\frac{5}{14}} \left\{4 |A_0|^2 \sin ^2\theta
   _1+|A_1|^2 \left[\cos \left(2 \theta
   _1\right)+3\right]\right\}\nonumber.
\end{eqnarray}

\begin{eqnarray}
\mathcal  A^0_0&=&2|B_1|^2+|B_0|^2,\nonumber\\
\mathcal  A^0_0\mathcal  A^4_2&=&\frac{1}{2} \sqrt{\frac{5}{14}} \left(3 |B_0|^2-4 |B_1|^2\right)
   \sin ^2\theta _2 \cos \left(2 \phi _2\right)\nonumber\\
   &\times& \left[7 \cos \left(2 \theta _2\right)+5\right],\nonumber\\
\mathcal  A^0_0\mathcal  A^4_1&=&\frac{1}{4} \sqrt{\frac{5}{7}} \left(3| B_0|^2-4 |B_1|^2\right)
   \sin \theta _2 \cos\phi _2\nonumber\\
   &\times&\left[9 \cos \left(\theta _2\right)+7 \cos \left(3\theta _2\right)\right],\nonumber\\
\mathcal  A^0_0\mathcal  A^4_0&=&\frac{\left(3 |B_0|^2-4 |B_1|^2\right) \left[20 \cos \left(2
   \theta _2\right)+35 \cos \left(4 \theta _2\right)+9\right]}{16 \sqrt{14}},\nonumber\\
\mathcal  A^0_0\mathcal  A^2_2&=&-\sqrt{\frac{30}{7}} \left(|B_0|^2+|B_1|^2\right) \sin^2\theta_2 \cos \left(2 \phi _2\right),\nonumber\\
\mathcal  A^0_0\mathcal  A^2_1&=&-\sqrt{\frac{30}{7}} \left(|B_0|^2+|B_1|^2\right) \sin \left(2
   \theta _2\right) \cos\phi _2 ,\nonumber\\
\mathcal  A^0_0\mathcal  A^2_0&=&-\sqrt{\frac{5}{14}} \left(|B_0|^2+|B_1|^2\right) \left[3 \cos
   \left(2 \theta _2\right)+1\right]\nonumber.
\end{eqnarray}

\section{$J^P=3^+$}
There are 10 multipole parameters ($r^L_M$) and analyzing powers ($\mathcal  A^L_M$) for the  $\zcs$ $J^P=3^+$ hypothesis. The definitions are referred to Table \ref{rlist}, and Eq. \ref{jad}. $A_{\lambda_1}$ and $B_{\lambda_2}$ are helicity amplitudes as defined in Table \ref{tab::angdef}.
\begin{eqnarray}
r^0_0&=&\frac{1}{4} \left\{2 |A_0|^2 \sin ^2\theta _1 +|A_1|^2 \left[\cos \left(2 \theta _1\right)+3\right]\right\},\nonumber\\
r^0_0r^6_2&=&\frac{1}{4} \sqrt{\frac{35}{33}} |A_1|^2 \sin ^2\theta _1,\nonumber\\
r^0_0r^6_1&=&-\frac{5}{12} \sqrt{\frac{7}{11}} \text{Re}(A_1 A^*_0)
   \sin \left(2 \theta _1\right),\nonumber
\end{eqnarray}
\begin{eqnarray}
r^0_0r^6_0&=&\frac{5 \left\{3 |A_1 |^2 \left[\cos \left(2 \theta _1\right)+3\right]-8 |A_0|^2 \sin \left(\theta _1\right){}^2\right\}}{24 \sqrt{22}},\nonumber\\
r^0_0r^4_2&=&-\frac{1}{2} \sqrt{\frac{5}{33}}| A_1|^2 \sin^2\theta _1,\nonumber\\
r^0_0r^4_1&=&\frac{1}{4} \sqrt{\frac{5}{11}} \text{Re}(A_1 A^*_0) \sin
   \left(2 \theta _1\right),\nonumber
\end{eqnarray}
\begin{eqnarray}
r^0_0r^4_0&=&\frac{12 |A_0|^2  \sin \left(\theta _1\right){}^2+|A_1 |^2
   \left(\cos \left(2 \theta _1\right)+3\right)}{8 \sqrt{33}},\nonumber\\
r^0_0r^2_2&=&\frac{1}{2 \sqrt{6}}|A_1|^2  \sin ^2\theta _1,\nonumber\\
r^0_0r^2_1&=&-\frac{1}{12} \text{Re}(A_1A^*_0 ) \sin \left(2 \theta
   _1\right),\nonumber\\
r^0_0r^2_0&=&\frac{-8 |A_0|^2  \sin \left(\theta _1\right){}^2-3 |A_1|^2
   \left[\cos \left(2 \theta _1\right)+3\right]}{24 \sqrt{2}}.\nonumber
\end{eqnarray}

\begin{eqnarray}
\mathcal  A^0_0&=&2|B_1|^2+|B_0|^2,\nonumber\\
\mathcal  A^0_0\mathcal  A^6_{2}&=&-\frac{5}{128} \sqrt{\frac{105}{11}} \left(2 |B_0|^2 -3 |B_1|^2\right) \sin^2\theta _2  \left\{60 \cos \left(2 \theta
   _2\right)\right.\nonumber\\
&+&\left.33 \cos \left(4 \theta _2\right)+35\right\} \cos \left(2 \phi
   _2\right),\nonumber\\
\mathcal  A^0_0\mathcal  A^6_1&=&-\frac{5}{128} \sqrt{\frac{21}{22}} \left(2 |B_0|^2 -3| B_1|^2 \right) \left\{5 \sin \left(2 \theta _2\right)\right.\nonumber\\
&+&\left.12 \sin \left(4 \theta
   _2\right)+33 \sin \left(6 \theta _2\right)\right\} \cos \left(\phi _2\right),\nonumber
\end{eqnarray}
\begin{eqnarray}
\mathcal  A^0_0\mathcal  A^6_0&=&-\frac{5}{256 \sqrt{22}} \left(2 |B_0|^2 -3 |B_1|^2 \right) \left\{105 \cos \left(2
   \theta _2\right)\right.\nonumber\\
   &+&\left.126 \cos \left(4 \theta _2\right)+231 \cos \left(6 \theta
   _2\right)+50\right\},\nonumber\\
\mathcal  A^0_0\mathcal  A^4_2&=&\frac{1}{4} \sqrt{\frac{15}{11}} \left(3 |B_0|^2 +|B_1|^2 \right)
   \sin ^2\theta _2\nonumber\\
   &\times&  \left[7 \cos \left(2 \theta _2\right)+5\right]
   \cos \left(2 \phi _2\right),\nonumber\\
\mathcal  A^0_0\mathcal  A^4_1&=&\frac{1}{8} \sqrt{\frac{15}{22}} \left(3 |B_0|^2 +|B_1 |^2\right)
   \left[2 \sin \left(2 \theta _2\right)+7 \sin \left(4 \theta _2\right)\right]\nonumber\\
   &\times& \cos \left(\phi _2\right),\nonumber\\
\mathcal  A^0_0\mathcal  A^4_0&=&\frac{1}{32} \sqrt{\frac{3}{11}} \left(3 |B_0|^2 +|B_1|^2 \right)\nonumber\\
   &\times&\left[20 \cos \left(2 \theta _2\right)+35 \cos \left(4 \theta
   _2\right)+9\right],\nonumber\\
\mathcal  A^0_0\mathcal  A^2_2&=&-\sqrt{\frac{3}{2}} \left(2 |B_0|^2 +3 |B_1|^2 \right) \sin
   \left(\theta _2\right){}^2 \cos \left(2 \phi _2\right),\nonumber\\
\mathcal  A^0_0\mathcal  A^2_1&=&-\sqrt{6} \left(2 |B_0|^2 +3 |B_1 |^2 \right) \sin \left(\theta
   _2\right) \cos \left(\theta _2\right) \cos \left(\phi _2\right),\nonumber\\
\mathcal  A^0_0\mathcal  A^2_0&=&-\frac{\left(2 |B_0|^2 +3 |B_1|^2 \right) \left(3 \cos \left(2
   \theta _2\right)+1\right)}{2 \sqrt{2}}.\nonumber
\end{eqnarray}

\section{$J^P=3^-$}\label{app::3m}
There are 7 multipole parameters ($r^L_M$) and analyzing powers ($\mathcal  A^L_M$) for the  $\zcs$ $J^P=3^-$ hypothesis. The definitions are referred to Table \ref{rlist}, and Eq. \ref{jad}. $A_{\lambda_1}$ and $B_{\lambda_2}$ are helicity amplitudes as defined in Table \ref{tab::angdef}.
\begin{eqnarray}
r^0_0&=&\frac{1}{4}| A_1 |^2 \left[\cos \left(2 \theta _1\right)+3\right],\nonumber\\
r^0_0r^6_2&=&-\frac{1}{4} \sqrt{\frac{35}{33}}| A_1|^2 \sin^2\theta _1,\nonumber\\
r^0_0r^6_0&=&\frac{5 |A_1|^2 \left[\cos \left(2 \theta _1\right)+3\right]}{8 \sqrt{22}},\nonumber
\end{eqnarray}
\begin{eqnarray}
r^0_0r^4_2&=&\frac{1}{2} \sqrt{\frac{5}{33}}|A_1|^2 \sin ^2\theta _1,\nonumber\\
r^0_0r^4_0&=&\frac{|A_1|^2 \left(\cos \left(2 \theta _1\right)+3\right)}{8 \sqrt{33}},\nonumber\\
r^0_0r^2_2&=&-\frac{|A_1|^2 \sin^2\theta _1}{2 \sqrt{6}},\nonumber
\end{eqnarray}
\begin{eqnarray}
r^0_0r^2_0&=&-\frac{|A_1|^2 \left(\cos \left(2 \theta _1\right)+3\right)}{8 \sqrt{2}}.\nonumber
\end{eqnarray}

\begin{eqnarray}
\mathcal  A^0_0&=&2|B_1|^2,\nonumber\\
\mathcal  A^0_0\mathcal  A^6_2&=&\frac{15}{128} \sqrt{\frac{105}{11}} |B_1|^2 \sin ^2\theta
   _2 \left\{60 \cos \left(2 \theta _2\right)\right.\nonumber\\
   &+&\left.33 \cos \left(4 \theta
   _2\right)+35\right\} \cos \left(2 \phi _2\right),\nonumber
\end{eqnarray}
\begin{eqnarray}
\mathcal  A^0_0\mathcal  A^6_0&=&\frac{15 }{256 \sqrt{22}}|B_1|^2 \left\{105 \cos \left(2 \theta _2\right)\right.\nonumber\\
&+&\left.126 \cos \left(4
   \theta _2\right)+231 \cos \left(6 \theta _2\right)+50\right\},\nonumber\\
\mathcal  A^0_0\mathcal  A^4_2&=&\frac{1}{4} \sqrt{\frac{15}{11}} |B_1|^2 \sin ^2\theta _2
   \left[7 \cos \left(2 \theta _2\right)+5\right]\cos \left(2 \phi _2\right)\nonumber,\\
\mathcal  A^0_0\mathcal  A^4_0&=&\frac{1}{32} \sqrt{\frac{3}{11}} |B_1|^2 \left[20 \cos \left(2 \theta
   _2\right)+35 \cos \left(4 \theta _2\right)+9\right],\nonumber\\
\mathcal  A^0_0\mathcal  A^2_2&=&-3 \sqrt{\frac{3}{2}} |B_1|^2 \sin ^2\theta _2 \cos \left(2
   \phi _2\right),\nonumber\\
\mathcal  A^0_0\mathcal  A^2_0&=&-\frac{3 |B_1|^2 \left[3 \cos \left(2 \theta _2\right)+1\right]}{2
   \sqrt{2}}.\nonumber
\end{eqnarray}

\section{Angular distribution parameters}
\label{app::angparlist}
Here are the lists of $\alpha_L(L=2,4,6)$ values defined in Table \ref{table::analist}.
\begin{eqnarray}
\alpha_2&=&\left\{
\begin{array}{ll}
{|A_1|^2-2|A_0|^2\over |A_1|^2+2|A_0|^2},&\text{for $1^+$,}\nonumber\\
{|A_1|^2-2|A_0|^2\over |A_1|^2+2|A_0|^2},&\text{for $2^-$,}\nonumber\\
{3|A_1|^2-4|A_0|^2\over 3|A_1|^2+4|A_0|^2},&\text{for $3^+$,}\nonumber\\
1& \text{for $1^-, 2^+$ and $3^-$.}
\end{array}\right.
\end{eqnarray}

\begin{eqnarray}
\alpha_4&=&\left\{
\begin{array}{ll}
{2|A_1|^2+3|A_0|^2\over 2|A_1|^2-3|A_0|^2},&\text{for $2^-$,}\nonumber\\
{|A_1|^2-6|A_0|^2\over |A_1|^2+6|A_0|^2},&\text{for $3^-$,}\nonumber\\
{|A_1|^2-6|A_0|^2\over |A_1|^2+6|A_0|^2},&\text{for $3^+$,}\nonumber\\
1& \text{for $2^+$.}
\end{array}\right.
\end{eqnarray}

\begin{eqnarray}
\alpha_6&=&\left\{
\begin{array}{ll}
{3|A_1|^2+4|A_0|^2\over 3|A_1|^2-4|A_0|^2},&\text{for $3^+$,}\nonumber\\
1& \text{for $3^-$.}
\end{array}\right.
\end{eqnarray}

%\end{widetext}
\vspace{1cm}
\section{Propagator for different $J^P$ hypotheses}
We give the propagators for the spin $J=1,2,3$ here.
\begin{widetext}
\begin{eqnarray*}
BW_{1}(k,\mu,\nu) &=& {\tilde{g}_{\mu\nu}(k) \over k^2-M^2-iM\Gamma},\text{~with~}\tilde{g}_{\mu\nu}(k)=-g_{\mu\nu}+
\displaystyle{k_\mu ~ k_\nu \over k^2},\text{~for J=1~},\\
BW_{2}(k,\mu,\nu,\alpha,\beta) &=& {i\over k^2-M^2-iM\Gamma}{\displaystyle{\left[{\displaystyle{1 \over 2}}(
\tilde{g}_{\nu \alpha }(k)\tilde{g}_{\mu \beta }(k)+\tilde{g}_{\nu \beta }(k)
\tilde{g}_{\mu \alpha }(k))-{\displaystyle{1 \over 3}}(\tilde{g}_{\alpha
\beta }(k)\tilde{g}_{\mu \nu }(k))\right] }}
,\text{~for J=2~},\\
BW_{3}(k,\mu,\nu,\alpha,\beta,\gamma,\eta) &=& {i\over k^2-M^2-iM\Gamma}\left[\right.{\displaystyle{1 \over 6}}(\tilde{g}_{\alpha \beta }(k)\tilde{g}_{\nu \gamma }(k)\tilde{g}_{\mu \eta }(k)+\tilde{g}_{\alpha \gamma }(k)\tilde{g}_{
\nu \beta }(k)\tilde{g}_{\mu \eta }(k)+\tilde{g}_{\alpha \beta }(k)
\tilde{g}_{\nu \eta }(k)\tilde{g}_{\mu \gamma }(k)  \\
&+&\tilde{g}_{\alpha \eta }(k)\tilde{g}_{\nu \beta }(k)\tilde{g}_{\mu \gamma }(k)+\tilde{g}_{
\alpha \gamma }(k)\tilde{g}_{\nu\eta }(k)\tilde{g}_{\mu \beta }(k)+
\tilde{g}_{\alpha \eta }(k)\tilde{g}_{\nu \gamma }(k)\tilde{g}_{\mu \beta }(k
))+{\displaystyle{-1 \over 15}}(   \\
&&\tilde{g}_{\mu \beta }(k)\tilde{g}_{
\gamma \eta }(k)\tilde{g}_{\nu \alpha }(k)+\tilde{g}_{\mu ,\gamma }(k)
\tilde{g}_{\beta \eta }(k)\tilde{g}_{\nu \alpha }(k)+\tilde{g}_{\mu ,\eta }(k)
\tilde{g}_{\beta \gamma }(k)\tilde{g}_{\nu \alpha }(k)+\tilde{g}_{\nu ,\beta
}(k)\tilde{g}_{\gamma \eta }(k)\tilde{g}_{\mu \alpha }(k)   \\
&+&\tilde{g}_{
\nu \gamma }(k)\tilde{g}_{\beta \eta }(k)\tilde{g}_{\mu ,\alpha }(k)+
\tilde{g}_{\nu \eta }(k)\tilde{g}_{\beta \gamma }(k)\tilde{g}_{\mu \alpha }(k
)+\tilde{g}_{\alpha \beta }(k)\tilde{g}_{\gamma \eta }(k)\tilde{g}_{\mu \nu
}(k)+\tilde{g}_{\alpha \gamma }(k)\tilde{g}_{\beta \eta }(k)\tilde{g}_{\mu
\nu }(k)   \\
&+&\tilde{g}_{\alpha \eta }(k)\tilde{g}_{\beta \gamma }(k)
\tilde{g}_{\mu\nu }(k))\left.\right],\text{~for J=3,~}
\end{eqnarray*}
\end{widetext}
where $k$ is the momentum of $D_sD^*$ system, and $M(\Gamma)$ is the mass(width) of $\zcs$.


\newpage
\begin{thebibliography}{99}
\bibitem{ali}Ahmed Ali, Jens S\"{o}ren Lange and Sheldon Stone,
Prog. Part. Nucl. Phys. {\bf97}, 123 (2017).
\bibitem{olsen}Stephen Lars Olsen, Tomasz Skwarnicki and  Daria Zieminska,
 Rev. Mod. Phys. {\bf90}, 015003 (2018).
\bibitem{yuancz}Nora Brambilla, Simon Eidelman, Christoph Hanhart, et.al,
Phys. Rept. {\bf873}, 1 (2020).
\bibitem{zc3900} M. Ablikim, et.al. (BESIII Collaboration), Phys. Rev. Lett. {\bf110}, 252001 (2013).
\bibitem{zc4020} M. Ablikim, et.al. (BESIII Collaboration), Phys. Rev. Lett. {\bf111}, 242001 (2013).
\bibitem{beszc2} M. Ablikim et al. (BESIII Collaboration), Phys. Rev. Lett. {\bf112}, 022001 (2014).
\bibitem{beszc3} M. Ablikim et al. (BESIII Collaboration), Phys. Rev. Lett. {\bf112}, 132001 (2014).
\bibitem{beszc4} M. Ablikim et al. (BESIII Collaboration), Phys. Rev. Lett. {\bf113}, 212002 (2014).
\bibitem{beszc5} M. Ablikim et al. (BESIII Collaboration), Phys. Rev. Lett. {\bf115}, 112003 (2015).
\bibitem{beszc6} M. Ablikim et al. (BESIII Collaboration), Phys. Rev. Lett. {\bf115}, 182002 (2015).
\bibitem{beszc7} M. Ablikim et al. (BESIII Collaboration), Phys. Rev. Lett. {\bf115}, 222002 (2015).
\bibitem{beszc8} Z. Q. Liu et al. (Belle Collaboration), Phys. Rev. Lett. {\bf110}, 252002 (2013);
 Erratum: [Phys. Rev. Lett. {\bf111}, 019901 (2013)].
\bibitem{beszc9} T. Xiao, S. Dobbs, A. Tomaradze and K. K. Seth, Phys. Lett. B {\bf727}, 366 (2013).
\bibitem{beszcs} M. Ablikim, et.al, (BESIII Collaboration), Phys. Rev. Lett., {\bf126},102001 (2020).
\bibitem{theozcs} M. B. Voloshin, Phys. Lett. B {\bf798}, 135022 (2019);\\
S. H. Lee, M. Nielsen and U. Wiedner, J. Korean Phys.
Soc. {\bf55}, 424 (2009);\\
J. Ferretti and E. Santopinto, JHEP {\bf04}, 119 (2020);\\
J. M. Dias, X. Liu and M. Nielsen, Phys. Rev. D {\bf88}, 096014 (2013);\\
D. Y. Chen, X. Liu and T. Matsuki, Phys. Rev. Lett. {\bf110}, 232001 (2013).
\bibitem{wangqn}Qi-Nan Wang, Wei Chen and Hua-Xing Chen, arXiv:2011.10495v2 [hep-ph].
\bibitem{ohlsen}G. G. Ohlsen, Rep. Prog. Phys. {\bf35}, 717 (1972);
Hong Chen, and Rong-Gang Ping, Phys. Rev. D{\bf102}, 016021 (2020).
\bibitem{qmatrix}M. G. Doncel, P. Mery, L. Michel, P. Minnaert, and K. C. Wali, Phys. Rev. {\bf7}, 815 (1973).
\bibitem{eleader}C. Bourrely, J. Soffer, and E. Leader, Phys. Rep. {\bf59}, 95 (1980).
\end{thebibliography}
\end{document}